\documentclass[aps,prl,reprint]{revtex4-1}
\usepackage{amsmath,amssymb}
\usepackage[dvips]{graphicx}
\usepackage{bm}

\usepackage[usenames]{color} 

\begin{document}

\title{Metallic Behavior of Cyclotron Relaxation Time in Two-Dimensional Systems}

\author{Ryuichi Masutomi,$^1$ Kohei Sasaki,$^1$ Ippei Yasuda,$^1$ Akihito Sekine,$^1$\\ 
Kentarou Sawano,$^2$ Yasuhiro Shiraki,$^2$ and Tohru Okamoto$^1$}
\affiliation{$^1$Department of Physics, University of Tokyo, 7-3-1, Hongo, Bunkyo-ku, Tokyo 113-0033, Japan\\
$^2$Research Center for Silicon Nano-Science, Advanced Research Laboratories, Tokyo City University,
8-15-1 Todoroki, Setagaya-ku, Tokyo 158-0082, Japan}

\date{\today}

\begin{abstract}

Cyclotron resonance of two-dimensional electrons is studied
at low temperatures down to 0.4~K
for a high-mobility Si/SiGe quantum well
which exhibits a metallic temperature dependence of dc resistivity $\rho$.
The relaxation time $\tau_{\rm CR}$ shows a negative temperature dependence,
which is similar to that of the transport scattering time $\tau_t$ obtained
from $\rho$.
The ratio $\tau_{\rm CR}/\tau_t$ at 0.4~K increases
as the electron density $N_s$ decreases,
and exceeds unity when $N_s$ approaches the critical density
for the metal-insulator transition.
 
\end{abstract}

\maketitle

The metallic behavior in two-dimensional (2D) systems has attracted much attention \cite{Abrahams2001,Sarma2005,Spivak2010}.
In 1994, Kravchenko \textit{et al.} observed a strong metallic temperature dependence of 
the zero-magnetic-field ($B=0$) resistivity ($d\rho/dT>0$)
of high-mobility 2D electron systems (2DESs)
in Si metal-oxide-semiconductor field-effect transistors (Si-MOSFETs),
and a metal-insulator transition (MIT) 
by changing the electron density $N_s$ \cite{Kravchenko1994,Cham1980}.
According to the scaling theory of localization \cite{Abrahams1979},
which neglects electron-electron (\textit{e-e}) interactions,
there can be no metallic state in 2D at $B=0$
and the system should become an insulator in the zero-temperature limit.
Thus the observation of the 2D metallic behavior and the MIT
has caused much controversy and is still a widely debated subject 
\cite{Abrahams2001,Sarma2005,Spivak2010}.
Recently, theoretical calculations using the renormalization group equations suggest that the metallic phase is stabilized by
\textit{e-e} interactions even at $T=0$ \cite{Punnoose2005}. 
Because the experimental data on Si-MOSFETs are in agreement with this theory,  
Anissimova \textit{et al.} argue that the 2D MIT is an interaction driven 
and $N_s$-tuned quantum phase transition (QPT) \cite{Anissimova2007}.
From another point of view,
$\rho(T)$ was calculated using the finite temperature Drude-Boltzmann theory
by Das Sarma and Hwang
\cite{Sarma2005,DasSarmaandHwang,Hwang2005}.
The experimental data were quantitatively reproduced for different 2D systems
by taking into account the temperature dependent screening of residual impurities.
An entirely different explanation was given by Spivak and Kivelson,
who introduced random microemulsion phases between a Fermi liquid
and a Wigner crystal (WC) phase \cite{Spivak2010,Spivak}.
The reduction of $\rho$ with decreasing temperature was explained
in terms of a decrease in the WC fraction having higher spin entropy.

The metallic temperature dependence of $\rho$ has been observed for various low-density
2D electron and hole systems, such as Si-MOSFETs \cite{Kravchenko1994},
\textit{p}-SiGe quantum wells (QWs) \cite{Lam1997,Coleridge1997},
\textit{p}-GaAs/AlGaAs heterojunctions \cite{Hanein1998p,Simmons1998},
\textit{n}-AlAs QWs \cite{Papadakis1998},
\textit{n}-GaAs/AlGaAs heterojunctions \cite{Hanein1998n,Lilly2003} and
\textit{n}-Si QWs \cite{Okamoto2004,Lai2005}.
A common feature of these systems is a strong \textit{e-e} interaction,
which is characterized by the Wigner-Seitz radius $r_s \equiv (\pi N_s)^{-1/2}/a_B
=\pi^{1/2}(e/h)^2(m^{\ast}/\kappa\epsilon_0)N_s^{-1/2}$.
Here $a_B$ is the effective Bohr radius, $m^\ast$ is the effective mass,
and $\kappa$ is the dielectric constant.
It is also suggested that the spin and/or valley degeneracies play
an important role in the appearance of the metallic behavior
\cite{Okamoto2004,Okamoto1999,Gunawan2007}.

According to the Drude formula,
the $B=0$ dc resistivity is related to 
the transport scattering time $\tau_t$
as $\rho^{-1}=e^2 N_s \tau_t/m^\ast$.
Since $\tau_t$ is not sensitive to small-angle scattering events,
it is not identical to the single-particle relaxation time $\tau_s$ \cite{DasSarma1985}.
If $W_{k,k'}$ is proportional to 
the probability of scattering from state $k$ to $k'$
and $\theta$ is the scattering angle,
$\tau_t$ and $\tau_s$ are given by
\begin{eqnarray}
\frac{1}{\tau_t}&=&\int W_{k,k'} (1-\cos \theta) dk',\\
\frac{1}{\tau_s}&=&\int W_{k,k'} dk'.
\end{eqnarray}
It is believed that the ratio is large ($\tau_{t}/\tau_{s}\gg 1$) for
long-range scattering potentials
and small ($\tau_{t}/\tau_{s} \approx 1$)
for short-range scattering potentials.
Experimentally, the single-particle relaxation time have been deduced
from the magnitude of the Shubnikov-de Haas oscillations ($\tau_{\rm SdH}$) 
\cite{Harrang1985,Coleridge1991} and 
from the cyclotron resonance ($\tau_{\rm CR}$) \cite{Linke1993,Syed2004}.
The ratio $\tau_{t}/\tau_{\rm SdH}$ was measured for various semiconductor 2D systems
in order to identify the main low-temperature scattering mechanisms.
However, it has been pointed out that 
$\tau_{\rm SdH}$ is severely affected by density inhomogeneities
especially in high mobility samples \cite{Syed2004}.
In contrast, $\tau_{\rm CR}$ is thought to be insensitive to 
density inhomogeneities.
Furthermore, $\tau_{\rm CR}$ can be measured even at high temperatures
where the Shubnikov-de Haas oscillations disappear.

We study $\tau_{\rm CR}$ in a high mobility Si 2DES
which exhibits the metallic $T$-dependence of $\rho$.
For $N_s$ between 0.74 and 1.93 $\times 10^{15}~\mathrm{m^{-2}}$,
$\tau_{\rm{CR}}$ monotonically increases with decreasing temperature
in the range from 7~K to 0.4~K.
The overall behavior of the $T$-dependence of $\tau_{\rm{CR}}$ is similar to that of $\tau_t$. 
The ratio $\tau_{\rm CR}/\tau_t$ at 0.4~K increases
with decreasing $N_s$,
and $\tau_{\rm{CR}}$ becomes larger than $\tau_t$
in the region near the MIT.

We used a Si/SiGe heterostructure with a 20-nm-thick strained Si QW
sandwiched between relaxed $\mathrm{Si}_{0.8}\mathrm{Ge}_{0.2}$ layers \cite{Yutani1996}.
The electrons are provided by a Sb-$\delta$-doped layer 20~nm above the channel
and $N_s$ can be controlled by varying bias voltage $V_{\rm BG}$ 
of a $p$-type Si(001) substrate 2.1~$\mu$m below the channel at 20~K.
The 2DES has a high mobility of 49 m$^2$/V~s
at $N_s = 2.36 \times 10^{15}~\mathrm{m^{-2}}$ ($V_{\rm BG}=0$ V) and $T = 0.4$ K.
Although the sample was mounted in a pumped ${}^3$He refrigerator,
a thermal insulation system enabled us to vary the sample temperature up to 10~K
on the condition that the base temperature 
of the carbon bolometer was kept constant at 0.35~K.
The schematic drawing of the equipment
is described elsewhere \cite{Masutomi2010}. 
The cyclotron resonance (CR) measurements were performed using
100~GHz millimeter-wave radiation.
The radiation power was kept low enough so that electron heating effects can be neglected.
In order to measure $\rho$ and $N_s$, the sample was fabricated in Hall bar geometry
and Ohmic contacts were made outside the irradiation area.

\begin{figure}[t!]
\includegraphics[width=.9\linewidth]{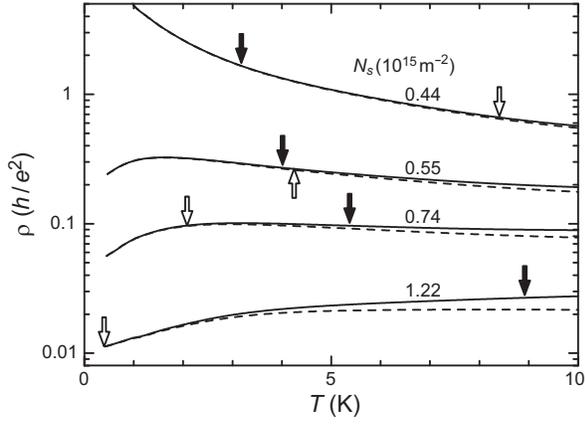}
\caption{
Temperature dependence of resistivity at $B=0$ for different electron densities.
The solid curves are the experimental data.
The dashed curves are obtained by subtracting
the contribution of phonon-scattering assuming 
Matthiessen's rule.
The black arrows indicate the Fermi temperature
and the white arrows indicate $T=\hbar/k_B \tau_t$.
}
\end{figure}

Figure 1 shows $T$-dependence of $\rho$ at $B=0$ for different $N_s$.
The dashed curves are obtained by subtracting
the contribution of phonon-scattering \cite{Paul1995} from the experimental data.
The Fermi temperature $T_F$ is indicated by black arrows.
Clear metallic $T$-dependence of $\rho$ can be seen
in the range $T \lesssim 0.3 T_F$ for 
$N_s > 0.5 \times 10^{15}~{\rm m}^{-2}$,
while it disappears for $N_s < 0.5 \times 10^{15}~{\rm m}^{-2}$.
The value of $r_s$ ranges from 7.5 to 5.0 for (0.55--$1.22) \times 10^{15}~{\rm m}^{-2}$,
indicating strong electron correlation in this system.
Furthermore the valley degeneracy of $g_v=2$ in Si 2DESs,
as well as the spin degeneracy of $g_s=2$,
reduces the Fermi energy ($\varepsilon_F=2\pi \hbar^2 N_s/g_v g_v m^\ast$)
and leads to the enhancement of the relative strength of the \textit{e-e} interaction.
The white arrows indicate the temperature 
at which the scattering rate $\tau_t{}^{-1}$ is equal to $k_B T /\hbar$.
Here, $\tau_t$ is related to the $B=0$ dc resistivity as $ \tau_t = m^\ast/e^2 N_s\rho $.
The metallic behavior can be observed both for
$k_B T< \hbar \tau_t{}^{-1}$ (low-$N_s$), where electrons propagate diffusively, and
for $k_B T> \hbar \tau_t{}^{-1}$ (high-$N_s$), referred to as the ballistic regime \cite{Zala2001}.

\begin{figure}[t!]
\includegraphics[width=.9\linewidth]{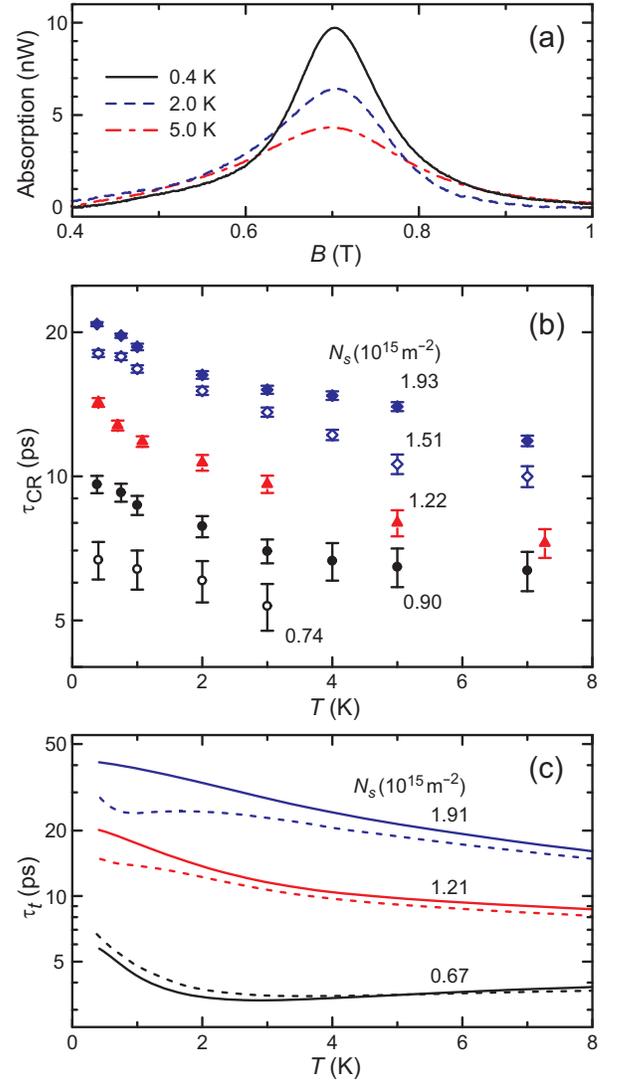}
\caption{(color online)
(a) CR traces at $N_s = 1.51 \times 10^{15}~{\rm m}^{-2}$ for $T=0.4$, 2.0 and 5.0~K.
(b) $T$-dependence of $\tau_{\rm{CR}}$ for $N_s=1.93$, 1.51, 1.22, 0.90,
$0.74 \times 10^{15}~{\rm m}^{-2}$ (from top to bottom).
(c) Solid curves are $T$-dependence of $\tau_t$ for $N_s=1.91$, 1.21,
$0.67 \times 10^{15}~{\rm m}^{-2}$ (from top to bottom).
The scattering time obtained from the longitudinal resistivity 
at $B=B_{\rm CR}$ is also plotted (dotted curves).
}
\end{figure}
Figure 2(a) shows typical CR observed at $N_s = 1.51 \times 10^{15}~{\rm m}^{-2}$
for different temperatures.
From the half-width at half maximum $\Delta B$, $\tau_{\rm CR}$ is obtained as
$\tau_{\rm{CR}} = B_{\rm CR}/(\omega\Delta B)$.
Here $B_{\rm CR}$ is the resonance magnetic field and $\omega$ is the microwave frequency
($\omega/2\pi=100$~GHz).
In Fig.~2(b), $\tau_{\rm CR}$ is plotted as a function of $T$ for different $N_s$.
In the wide density range, $\tau_{\rm CR}$ is metallic, i.e.,
it increases with decreasing temperature.
For comparison, $\tau_t$ determined from the $B=0$ dc resistivity
is shown in Fig.~2(c) (solid curves).
The $T$-dependence of $\tau_{\rm CR}$ is similar to that of $\tau_t$.
In order to study the effect of the magnetic field on the transport scattering time,
we have measured the longitudinal resistivity $\rho_{xx}$ at $B=B_{\rm CR}$.
According to the classical Drude picture,
the scattering time $\tau_b=m^\ast/e^2 N_s \rho_{xx}$ is obtained as a function of $T$
and shown in Fig.~2(c) as the dotted curves.
Except in the low-$T$ region for high $N_s$
where the Shubnikov-de Haas oscillations are superimposed,
the deviation of $\tau_b (T)$ from $\tau_t(T)$ is small.
This demonstrates that the magnetic field applied for the CR measurements
is small enough not to affect the transport scattering time significantly.

The results shown in Fig.~2 indicate that
the scattering time has the metallic $T$-dependence
over a very wide frequency range from dc to 100~GHz.
At 100~GHz, a characteristic length $l_\omega = v_F/\omega$
becomes much shorter ($l_\omega = 50 \sim 70$~nm) than the electron mean free path
$\lambda = 120 \sim 1900$~nm, where $v_F$ is the Fermi velocity.
Furthermore, the photon energy $\hbar \omega = 4.8$~K
exceeds the typical temperatures where
the metallic behavior is observed clearly.
At this stage, it is unclear why the metallic $T$-dependence of the scattering time
occurs under very different conditions.
However, we believe that our observations will provide a strong constraint on theoretical models.

In order to discuss the 2D metallic behavior further, it is important to identify
the type of disorder potential. 
In Figs.~3(a) and 3(b),
$\tau_{\rm CR}$, $\tau_t$ and $\tau_{\rm{CR}}/\tau_t$ at $T=0.4$~K are shown
as a function of $N_s$.
For high $N_s$ between 1 and 2 $\times 10^{15}~{\rm m}^{-2}$,
$\tau_{\rm{CR}}$ is smaller than but comparable to $\tau_t$.
This suggests that
short-range potential fluctuations which lead to large-angle scattering
play a dominant role in the metallic regime,
although the Si/SiGe 2DES studied is a high-mobility modulation-doped-heterostructure.
This is consistent with the recent work of Clarke \textit{et al.}
who compared the strength of the metallic behavior 
in GaAs 2D hole systems dominated by long- and short-range disorder \cite{Clarke2008}.

\begin{figure}[t!]
\includegraphics[width=1.0\linewidth]{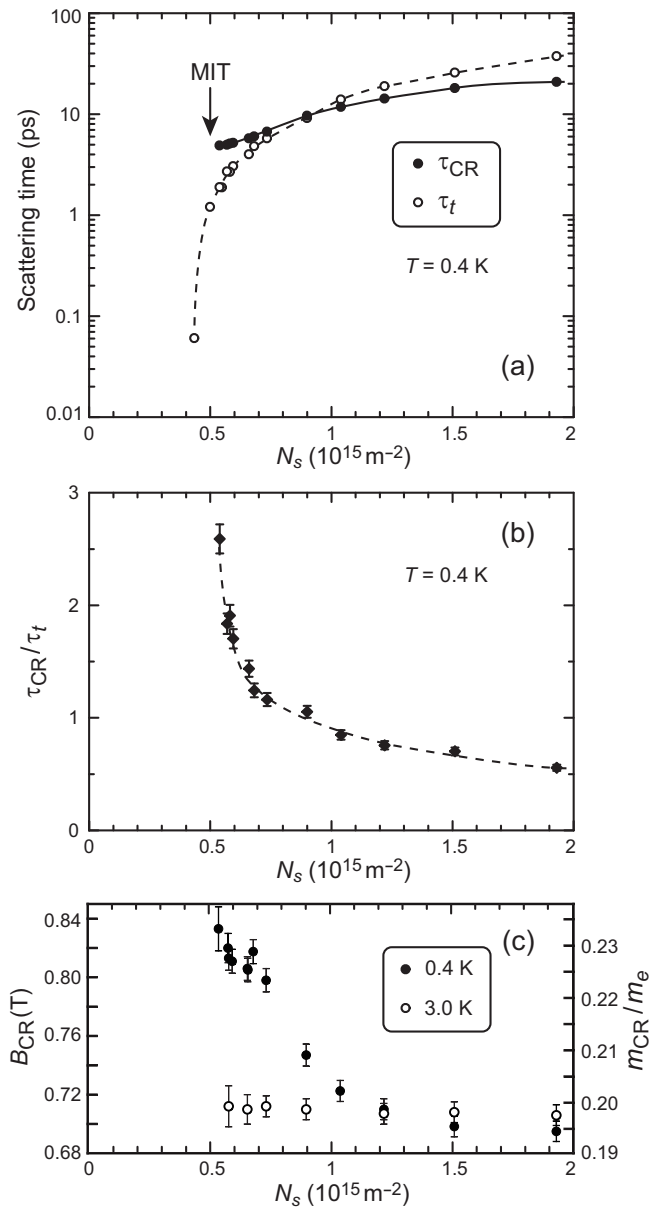}
\caption{
(a) Density dependence of $\tau_{\rm{CR}}$ (solid circle) and $\tau_t$ (open circle) at 0.4~K.
The arrow indicates the critical density for the MIT 
determined from the $T$-dependence of $\tau_t$.
Solid and dashed lines are a guide to the eye.
(b) The ratio of $\tau_{\rm{CR}}$ to $\tau_t$ at 0.4~K.
(c) The resonance magnetic field $B_{\rm{CR}}$ and the corresponding
effective mass $m_{\rm{CR}}$
at 0.4 K (solid circle) and 3.0 K (open circle).
}
\end{figure}

As $N_s$ decreases and approaches the critical density for the MIT,
$\tau_t$ rapidly decreases as shown in Fig. 3(a).
On the other hand, the $N_s$-dependence of $\tau_{\rm{CR}}$
seems to be gradual,
although we were not able to determine it for $N_s <0.5 \times 10^{15}~{\rm m}^{-2}$
due to the broadening of the CR signal.
The ratio $\tau_{\rm{CR}}/\tau_t$ at $T=0.4$~K exceeds unity near the MIT as shown in Fig. 3(b).
This cannot be understood in a single-particle picture on which Eqs.~(1) and (2) is based.
In Fig. 3(c), $B_{\rm CR}$ and the corresponding effective mass
$m_{\rm CR}=eB_{\rm CR}/\omega$ are shown as a function of $N_s$.
While $B_{\rm CR}$ is independent of $N_s$ and 
$m_{\rm CR}$ is close to $m^\ast=0.19m_e$ at $T=3.0$~K,
$B_{\rm CR}$ increases rapidly with
decreasing $N_s$ near the MIT at $T=0.4$~K.
In Ref.~[\onlinecite{Wilson1980}],
a narrowing of the CR absorption line,
together with a deviation of the resonance frequency from $\omega_c=eB/m^\ast$,
has been reported for a Si-MOSFET
in the extreme quantum limit.
The observed width is about 5 times narrower than 
that expected from $\tau_t$.
The results were explained in terms of the formation of
a magnetic-field-induced Wigner glass (WG).
Even at $B=0$, a pinned WC or WG is expected to be formed
in low-density 2D semiconductor systems.
Pudalov \textit{et al.} observed nonlinear dc conduction
with a sharp threshold electric field in the insulating regime
of Si-MOSFETs and attributed it to that of a pinned WC \cite{Pudalov1993}.
Chui and Tanatar found from their Monte Carlo studies that the WC
can be stabilized at $r_s= 7.5$ in the presence of a very
small amount of disorder \cite{Chui1995}.
Our results for low $N_s$ imply that
electron correlation effects are important even in the metallic region at low temperatures.

In summary, we have performed the cyclotron resonance measurements
on a high-mobility Si 2DES.
The relaxation time $\tau_{\rm CR}$, obtained from the linewidth,
was found to have a negative $T$-dependence,
which is similar to that of $\tau_t$
corresponding to the metallic dc resistivity.
In the region around the MIT,
$\tau_{\rm CR}$ exceeds $\tau_t$ and
$B_{\rm CR}$ becomes larger than $m^\ast \omega /e$.
These unexpected behaviors cannot be described as a non-interacting electron system.
Further theoretical calculations taking into account cyclotron resonance in the metallic phase are eagerly desired.

The authors thank M. Koshino for helpful discussions.
This work has partly supported by Grant-in-Aid for Scientific Research (B) (No. 18340080), (A) (No. 21244047), and Grant-in-Aid for Scientific Research on Priority Area ``Physics of New Quantum Phases in Superclean Materials'' (No. 20029005) from MEXT, Japan.

\end{document}